# Response to Comment on "Spin-Orbit Logic with Magnetoelectric Nodes: A Scalable Charge Mediated Nonvolatile Spintronic Logic" (ArXiv: 1607.06690)

Sasikanth Manipatruni, Dmitri E. Nikonov, Huichu Liu* and Ian A. Young

*Components Research, Intel Corp., Hillsboro, OR 97124, USA*

*Intel Labs, Intel Corp., Santa Clara, CA 95054, USA*

In this technical note, we address the comments on the energy estimates for Magnetoelectric Spin-orbit (MESO) Logic, a new logic device proposed by the authors. We provide an analytical derivation of the switching energy, and support it with time-domain circuit simulations using a self-consistent ferroelectric (FE) compact model. While the energy to charge a capacitor is dissipated in the interconnect and transistor resistance, we note that the energy to switch a capacitor and a FE is independent of the interconnect resistance value to the first order. Also device design can mitigate the parasitic energy losses. We further show the circuit simulations for a sub 10 aJ switching operation of a MESO logic device comprehending: a) Energy stored in multiferroic; b) Energy dissipation in the resistance of the interconnect, $R_{ic}$; c) Energy dissipation in the inverse spin-orbit coupling (ISOC) spin to charge converter $R_{isoc}$; d) Supply, ground resistance, and transistor losses. We also identify the requirements for the resistivity of the spin-orbit coupling materials and address the effect of internal resistance of the spin to charge conversion layer. We provide the material parameter space where MESO (with a fan-out of 1 and interconnect) achieves sub 10 aJ switching energy with path for scaling via ferroelectric/magnetoelectric/spin-orbit materials development.

The switching energy of a Magnetoelectric Spin-Orbit (MESO) logic device [1] is composed of the sum of all the dissipation sources and energy storage: energy stored in the magnetoelectric ferroelectric (FE) ($E_{CME}$), and dissipation in the interconnect ($E_{IC}$), dissipation in the spin to charge conversion layer ($E_{ISOC}$), dissipation in the supply and ground path resistances ($E_{SG}$), and dissipation in the power supply transistor resistance ($E_{RT}$).

The total energy of a MESO state transition

$$E_{MESO} = E_{CME} + E_{IC} + E_{ISOC} + E_{RT} + E_{SG} \quad (1)$$

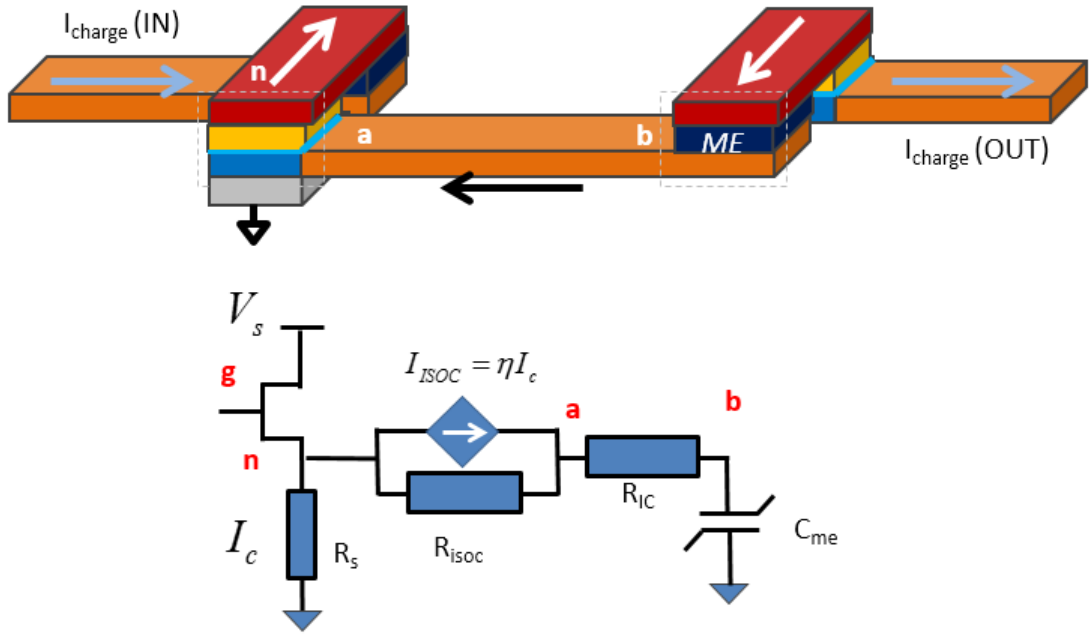

*Figure 1. **((Charge equivalent circuit for MESO))** MESO device [1] comprising a magnetoelectric (ME) switchable capacitor representing the MESO input loading combined with a spin to charge transduction mechanism for the MESO magnetic state read-out. The spin to charge conversion is modeled as a current controlled current source with an internal resistance $R_{isoc}$, the charge dynamics of the ME are modeled with a ferroelectric capacitor. $R_{ic}$ is the*

*interconnect resistance forming the charge interconnect ab. η is net spin to charge current conversion ratio.*

1. **Equivalent charge circuit model for MESO logic device for energy calculation**

MESO logic device comprises: a) magnetoelectric capacitor for voltage controlled switching of a ferromagnet (FM); b) a spin to charge conversion layer for charge readout of the magnetic state of the FM; and c) a charge interconnect connecting the MESO devices. An equivalent lumped element charge circuit model to capture the functioning of the MESO device is shown in Figure 1. Nodes *a* and *b* represent the two ends of the charge interconnect with interconnect resistance ($R_{ic}$). The spin to charge conversion element formed with inverse spin-orbit coupling (ISOC) materials is shown between node *n* and *a.* The ISOC module is modeled as a current controlled current source (CCCS) with an internal parallel resistance ($R_{isoc}$) [2, 3]. The charge dynamics of the magnetoelectric node are modeled via a ferroelectric capacitance ($C_{me}$) [4]. We present both an analytical expression for the total energy of a state transition of MESO, including the sum of all the dissipation and storage sources viz. energy stored in the ferroelectric (magnetoelectric), dissipation in the interconnect, dissipation in the supply-ground path, dissipation in the power supply transistor. We provide an analytical derivation of the transition energy, and support this with time-domain circuit simulations using a self-consistent ferroelectric compact model. The equivalent charge circuit is applicable for calculating the energy due to charge dynamics and does not capture the vector spin dynamics [1].

The explanation for the total energy of MESO device is as follows a) The energy to switch a capacitor and FE is independent of the interconnect resistance to the first order [2, 3] b) The current shunted in the spin to charge conversion current source depends on the equivalent source resistance ($R_{ISOC}$), which is material dependent parameter c) The losses in the parasitic paths are second order and device design can mitigate the energy losses extrinsic to magnetoelectric/ferroelectric switching. We also present an example method for mitigating energy losses extrinsic to magnetoelectric switching via proper choice of supply-ground path resistance ($R_s$). Section 2 provides a simple analytical derivation for the MESO switching energy. Section 3 provides detailed material and device parameters as well as analytical models used in benchmarking MESO with CMOS [5]. Section 4 describes results of time domain circuit simulation showing: a) the operating regime for MESO; b) impact of internal resistance of the ISOC source ($R_{isoc}$) and supply-ground path resistance ($R_s$). It provides an example of MESO operation at 10 aJ per switching transition comprehending all the energy delivered from the power supply. In Section 5, we respond to specific statements of the comment [6].

## 2. Energy calculation and scaling for MESO logic device

### a. Switching energy of a capacitor/ferroelectric is independent of the interconnect resistance

We first note that the energy to switch a capacitor (or a fixed charge switchable device such as a ferroelectric) is independent of the interconnect resistance. For simplicity, we start with a linear dielectric capacitor. In the first stage it is switched from voltage 0 to voltage V through an interconnect of resistance of R by a voltage supply of V. We see that, the interconnect dissipated energy is independent of the interconnect resistance,

$$E_{Ohmic} = R\int_0^\infty i^2 dt = \frac{V^2}{R}\int_0^\infty e^{-2t/RC} dt = \frac{CV^2}{2} \qquad (2.1)$$

The total energy supplied by the voltage source is given by

$$E_{supply} = V\int_0^\infty i\, dt = \frac{V^2}{R}\int_0^\infty e^{-t/RC} dt = CV^2 \qquad (2.2)$$

So one half of it dissipated in the charging stage, according to (2.1), and the other half goes to the increase of energy of the capacitor. In the second stage, the voltage supply is removed, and the capacitor evolves from voltage V to voltage 0. In this stage the energy stored in the capacitor is dissipated in the resistor. Therefore, the overall energy dissipated in the charge-discharge cycle is equal to (2.2).

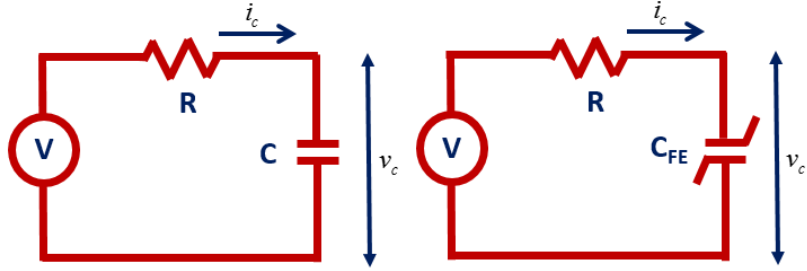

*Figure 2. RC circuit to show that the interconnect resistance does not impact the total energy to switch a capacitor. Linear dielectric capacitor used in the schematic on the left, a ferroelectric capacitor is used in the schematic on the right.*

The ferroelectric capacitor is treated somewhat differently. It starts with zero voltage but non-zero charge $-Q_{fe}$, corresponding to reversed spontaneous polarization $-P_{fe}$. It is then charged to voltage V and polarization $P_{fe}$. In the discharge stage, the voltage returns to zero, but there still remains the charge of $Q_{fe}$, corresponding to spontaneous polarization $P_{fe}$. The energy of the ferroelectric capacitor remains the same after the charge-discharge cycle. Similar to the linear dielectric capacitor, the total energy supplied by the voltage source to switch a ferroelectric is independent of the pulse shape of the current and is given by

$$E_{supply} = V\int_0^\infty idt = 2VQ_{fe} = C_{ME}V^2 \qquad (2.3)$$

where we introduced the ferroelectric capacitance $C_{me}$. Hence, the Ohmic losses in switching a ferroelectric capacitor are only different from a linear capacitor by the factor of 2.

## b. Switching energy of MESO

The total energy consumption of MESO can be written as sum of all the dissipation and storage sources Viz. energy stored in the FE, dissipation in the interconnect, dissipation in the supply-ground path, dissipation in the power supply transistor. We now consider two simplified equivalent models for the MESO for the energy consumption calculation comprising of the power supply (pulsed), Supply-ground path for spin to charge conversion, SOC current source, interconnect resistance and equivalent capacitance for ME. Figure 3A shows the simplified equivalent circuit to derive an analytical expression. We convert the spin to charge conversion current source (a current controlled current source) to a voltage source using Thevenin equivalence in Fig.3B.

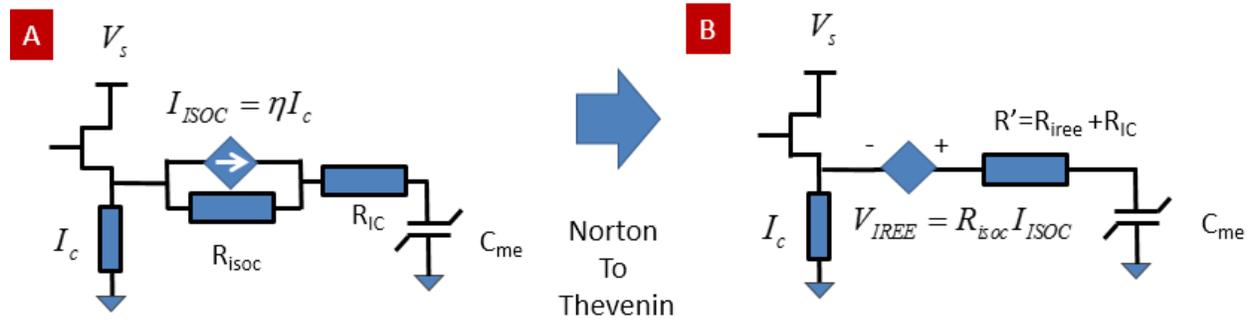

*Figure 3. A) Equivalent circuit for MESO with current controlled current source B) Current controlled voltage source. We apply Norton to Thevenin conversion.*

*Energy to switch the FE and the interconnect losses:* The total power dissipated in the interconnect resistance $R_{ic}$ and the ISOC internal resistance $R_{isoc}$ can be combined into R'. The power dissipated in the $R_{ic}$, $R_{isoc}$ and stored in ME capacitor are given by:

$$E_{CME} + E_{IC} + E_{ISOC} = C_{me} V_{me}^2 \qquad (2.5)$$

*Joule losses occur in the resistances in the current delivery paths of the transistor and the resistive supply-ground path:* The energy dissipated in the supply-ground path is given by

$$E_{SG} = \int_0^\infty i_{sh}^2 R_{sh} dt = i_{sh}^2 R_{sh} \frac{2Q}{I_c} = k C_{me} V_{me}^2 \frac{W}{\lambda} \qquad (2.6)$$

Where we use the relation, $i_{sh} R_{sh} = k V_{me}$, $i_{sh}/i_c = W/\lambda$, $2Q = C_{me} V_{me}$. W is the width of the magnet and $\lambda$ is the ISOC parameter. It can be further shown that, the energy dissipated in the transistor and supply-ground path is given by

$$E_{SG} + E_{RT} = \alpha C_{me} V_{me}^2 \frac{W}{\lambda} \qquad (2.7)$$

Where $\alpha$ is a circuit dependent function of transistor resistance, supply-ground path resistance. The total energy of MESO can be written as

$$E_{MESO} = E_{CME} + E_{IC} + E_{ISOC} + E_{RT} + E_{SG} = C_{me} V_{me}^2 \left(1 + \alpha \frac{W}{\lambda}\right) \qquad (2.8)$$

### 3. MESO parameters and benchmarks

Here we provide a detailed MESO switching energy calculation following the methodology of beyond-CMOS benchmarking [5]. In the course of derivation we will elucidate the misunderstanding about the MESO mode of operation leading to incorrect performance estimates [6]. As the reader will see, the estimates below are in approximate agreement with the rigorous SPICE simulations (Section 4). For our estimates we assume the following material and device parameters (Table 1). The calculation of the circuit area is outside the scope of this section. The charge circuit model of MESO operation is shown in Figure 1. Even though CMOS auxiliary circuits play a crucial role in the operation of MESO, here we will not describe the performance of a CMOS transistor. For details please see [5].

I. <u>Magnetoelectric effect.</u>

In order to achieve the magnetoelectric field necessary to switch a nanomagnet, the following electric field needs to be applied to the multiferroic BiFeO$_3$:

$$E_r = E_{mf} B_c / B_{mf}. \qquad (3.1)$$

(For the definition of terms in Eq. (3.1) see Table 1 below). The total charge at the terminals of the multiferroic capacitor comprises the saturated ferroelectric polarization charge at the interface and the linear dielectric polarization in response to the applied electric field:

Table 1. Material and structure parameters serving as inputs into MESO estimates.

| Quantity | Symbol | Units | Value |
|---|---|---|---|
| Characteristic critical dimension | $F$ | m | 1e-8 |
| Copper wire resistivity | $\rho_{Cu}$ | Ω*m | 2.5e-7 |
| Magnetization in a ferromagnet, perpendicular | $M_{sp}$ | A/m | 3e5 |
| Thickness of ferromagnetic | $t_{fm}$ | m | 2e-9 |
| Spin polarization from a ferromagnet | $P_{fm}$ |  | 0.7 |
| Perpendicular magnetic anisotropy | $K_u$ | J/m$^3$ | 6e5 |
| Spin-orbit coupling effect coefficient | $\lambda_{ISOC}$ | m | 1.4e-8 |
| Resistance*area of the FM and ISOC stack | $r_{fm}$ | Ω*m$^2$ | 3e-14 |
| Internal resistance of the ISOC current source | $R_{isoc}$ | Ω | 4000 |
| Magnetoelectric field for switching nanomagnet | $B_c$ | T | 0.1 |
| Multiferroic ferroelectric polarization (BFO) | $P_{mf}$ | C/m$^2$ | 0.3 |
| Multiferroic electric switching field | $E_{mf}$ | V/m | 1.8e6 |
| Multiferroic exchange bias at switching field | $B_{mf}$ | T | 0.03 |

| Dielectric constant of multiferroic | $\varepsilon_{mf}$ | | 54 |
|---|---|---|---|
| Thickness of multiferroic | $t_{mf}$ | m | 5e-9 |
| Ferroelectric intrinsic switching time | $\tau_{fe}$ | s | 2e-11 |
| Lande factor | $g$ | | 2 |
| Gate voltage for the access transistor | $V_x$ | V | 0.73 |
| On current for the access transistor | $\tilde{I}_x$ | A/m | 1648 |
| Gate capacitance per unit width of the access transistor | $c_g$ | F/m | 1e-9 |
| Resistance of a power or a ground distribution network | $R_s$ | Ω | 4000 |

$$Q_{me} = A_{me}\left(\varepsilon_0 \varepsilon_{mf} E_r + P_{mf}\right). \tag{3.2}$$

The voltage drop on the magnetoelectric element is

$$V_{me} = E_r t_{mf}. \tag{3.3}$$

The time to charge the multiferroic capacitor from 0 voltage to $V_{me}$ is

$$\tau_{me} \approx 2Q_{me} / I_{ISOC}, \tag{3.4}$$

Since the charge in the capacitor needs to be changed from $-Q_{me}$ to $Q_{me}$. Here $I_{ISOC}$ is the current produced by the spin-orbit effect.

From the treatment below we will see that the capacitor charging time is limited also by the intrinsic switching time for the ferroelectric BFO to reverse polarization, $\tau_{fe}$. However the magnetization is even slower to react to the applied exchange bias and takes optimistically the following time (optimistically) to complete the precession:

$$\tau_{mag} \approx \pi/(\gamma B_c). \tag{3.5}$$

Where the gyromagnetic ratio is

$$\gamma = ge/(2m_e). \tag{3.6}$$

The total time is obtained as the combination of the above two times:

$$\tau_{tot} = \tau_{me} + \tau_{mag}, \tag{3.7}$$

Where the last term makes the dominant contribution. The capacitance of the magnetoelectric element is

$$C_{me} = Q_{me}/V_{me}, \tag{3.8}$$

And the switching energy is

$$E_{me} = 2Q_{me}V_{me}. \tag{3.9}$$

    II.    <u>Spin-orbit coupling effect.</u>

The power supply enable transistor provides current $I_c$ traveling through the ferromagnet. This current is related to the supply voltage $V_{sup}$ and the total resistance in the supply path.

$$V_s = I_c(R_T + R_{fm} + R_s) \equiv I_c R_{sum}. \tag{3.10}$$

The enable transistor is in the linear regime. Its resistance is related to its width:

$$R_{ON} = r_L/w_x. \tag{3.11}$$

This transistor width ($w_x$) is set by the objective to provide sufficient current to be converted by spin-orbit effect. The value of this width turns out to be smaller than a minimal transistor width.

That means that one transistor can be shared to supply current to several parallel MESO devices, or a longer than minimum transistor channel length is used.

The resistance across the ferromagnet and spin-orbit coupling stack in the supply path is:

$$R_{fm} = r_{fm}/A_{me}. \tag{3.12}$$

We assume that the contact resistance is included in the definition of the I-V characteristic of the MOSFET. $R_{ON}$ resistance is related to the on-current per unit width at the low source-to drain voltage and high gate to source voltage

$$r_L \approx V_x/(3\tilde{I}_x). \tag{3.13}$$

The current extracts spin polarized current from the ferromagnet in the vertical direction

$$I_s = P_{fm}I_{sup}. \tag{3.14}$$

Inverse spin-orbit coupling effect (the combination of the bulk spin Hall effect and the interface Rashba-Edelstein effect) converts the spin polarized current into charge current in the charging path (horizontal direction) of the multiferroic capacitor of the next MESO gate with its sign determined by the direction of magnetization in the ferromagnet.

$$I_{ISOC} = \lambda_{ISOC}I_s/w_m, \tag{3.15}$$

Where $w_m$ is defined in table 1. This current source is related to the voltage it can produce, which must be equal to the magnetoelectric voltage:

$$V_{me} = I_{ISOC}(R_{ISOC} + R_{fe} + R_{ic}) = I_{ISOC}R_{tot}. \tag{3.16}$$

The resistances in the charging path in the equation above are the ISOC current source internal resistance (longitudinal resistance of the thin ISOC layer), the resistance of the ferroelectric capacitor, and the interconnect resistance. The resistance of the ferroelectric capacitor is caused by damping in the ferroelectric and is related to the ferroelectric intrinsic switching time

$$R_{fe} = \tau_{fe} / C_{me}. \tag{3.17}$$

The ISOC current source needs to be active over time $\tau_{me}$, necessary to charge the magnetoelectric capacitor. Thus energy dissipated in the vertical, supply path of the circuit is given by Joule power dissipation as:

Table 2. Operating parameters of MESO devices.

| Quantity | Symbol | Units | Value |
| --- | --- | --- | --- |
| Metal wire pitch (=4F) | $p_m$ | m | 4e-8 |
| Supply voltage | $V_s$ | V | 0.1 |
| Access transistor width | $w_x$ | m | 2e-9 |
| Resistance per width of the access transistor in the linear regime | $r_L$ | Ω*m | 1.48e-4 |
| Total resistance in the supply path | $R_{sum}$ | Ω | 2.5e4 |
| Characteristic interconnect length (=10p$_m$) | $l_{ic}$ | m | 4e-7 |
| Capacitance of a characteristic interconnect | $C_{ic}$ | F | 3.7e-17 |
| Resistance of a characteristic interconnect | $R_{ic}$ | Ω | 1e3 |
| Width of the magnet (=F) | $w_m$ | m | 1e-8 |
| Magnetoelectric switching area (=F²) | $A_{me}$ | m² | 1e-16 |
| Magnetoelectric voltage | $V_{me}$ | V | 0.03 |
| Switching time for the magnetoelectric capacitor | $\tau_{me}$ | s | 5e-11 |
| Effective magnetoelectric capacitance | $C_{me}$ | F | 1e-15 |
| Resistance of the ferroelectric capacitor | $R_{fe}$ | Ω | 1e4 |

| Total resistance in the charging path | $R_{tot}$ | Ω | 1.8e4 |
|---|---|---|---|
| Switching time for magnetization | $\tau_{mag}$ | s | 2e-10 |
| Current in the supply path | $I_c$ | A | 1.2e-6 |
| Current generated by the ISOC effect | $I_{ISOC}$ | A | 1.2e-6 |
| Energy of magnetoelectric capacitor | $E_{me}$ | J | 1.8e-18 |
| Energy dissipation in the supply path | $E_{sup}$ | J | 6.4e-18 |
| Energy to charge the access transistor | $E_{ga}$ | J | 1.1e-18 |

$$E_{sup} = V_s I_c \tau_{me}. \tag{3.18}$$

We can show that

$$E_{sup} = V_{sup} Q \frac{w_m}{P_{fm} \lambda_{IREE}}. \tag{3.19}$$

Thus this contribution into energy dissipation is related to the energy of the magnetoelectric capacitor too. Together the energy loss in the $C_{me}$ charging/discharging and supply paths is

$$E_{MESO} = E_{me}\left(1 + \frac{w_m}{P_{fm} \lambda_{IREE}} \frac{V_s}{V_{me}}\right). \tag{3.20}$$

An additional energy loss comes from charging the gate of the power supply gating transistor

$$E_{ga} = w_x c_g V_x^2. \tag{3.21}$$

### III. MESO switching performance.

For a characteristic interconnect between MESO elements, the switching time is calculated according [7]:

$$t_{ic} \approx 0.38 R_{ic} C_{ic} + 0.7 R_{on} C_{ic} + 0.7 R_{ic} C_L, \qquad (3.22)$$

as well as the interconnect switching energy:

$$E_{ic} \approx C_{ic} V_{me}^2. \qquad (3.23)$$

Here one needs to substitute for the load capacitance $C_L = C_{me}$ and for the on-state resistance $R_{on} = (R_{ISOC} + R_{ic})$.

The total delay time of an intrinsic device (not including the interconnect) is

$$t_{int} \approx \tau_{tot}, \qquad (3.24)$$

The intrinsic device in our case means 'one MESO element'. The way we calculate the performance of more complicated circuits, such as a 32-bit adder and an ALU, follows [5].

Energy dissipation in the charging path of the circuit, is not a separate contribution to switching energy. It is equal to the energy accumulated in the magnetoelectric capacitor and just represents the way this energy is dissipated each time the voltage is turned on or off (see Section II for the explanation).

Thus the total intrinsic device switching energy is composed of

$$E_{int} = E_{me} + E_{sup} + E_{ga}. \qquad (3.25)$$

The results of the calculation following the method of [7] are summarized in Table 3.

Table 3. Resulting performance of MESO devices and circuits.

| Quantity | Symbol | Units | Value |
|---|---|---|---|
| Area of the intrinsic device | $a_{int}$ | m$^2$ | 1.4e-14 |
| Switching time of the intrinsic device | $t_{int}$ | s | 2.3e-10 |
| Switching energy of the intrinsic device | $E_{int}$ | J | 9.3e-18 |
| Switching time of the interconnect | $t_{ic}$ | s | 2.9e-12 |
| Switching energy of the interconnect | $E_{ic}$ | J | 1.8e-19 |
| Area of 1 bit of a full adder | $a_1$ | m$^2$ | 8.6e-14 |
| Switching time of 1 bit of a full adder | $t_1$ | s | 2.4e-10 |
| Switching energy of 1 bit of a full adder | $E_1$ | J | 1.3e-16 |

Therefore we confirm the original estimate [1] that the switching energy of the MESO device with an interconnect is below 10aJ.

## 4. SPICE simulation of the charge circuit with equivalent ferroelectric model

We validated our assumptions of the charge transport in the MESO device via a SPICE circuit simulation solver using compact model that comprehends the physics of the ferro-electrics and the spin to charge conversion. We model the ferroelectric switching dynamics of the magnetoelectric using Landau-Khalatnikov (LK) equation:

$$\rho \frac{dQ_F}{dt} = -\frac{dU}{dQ_F} = -\left(2\alpha Q_F^1 + 4\beta Q_F^3 + 6\gamma Q_F^5 - V_F\right) \quad (4.1)$$

Where $Q_F$ is the ferroelectric polarization, $\rho$ internal equivalent resistance (damping term) of the ferroelectric, U is the energy density per unit area, α, β, γ are the rescaled internal anisotropy constants of the FE. The ferroelectric switching exhibits a non-linear equivalent capacitance

$$C_F(Q_F) = \left(2\alpha + 4\beta Q_F^2 + 6\gamma Q_F^4\right)^{-1} \qquad (4.2)$$

during the charging and discharging of the ferroelectric.

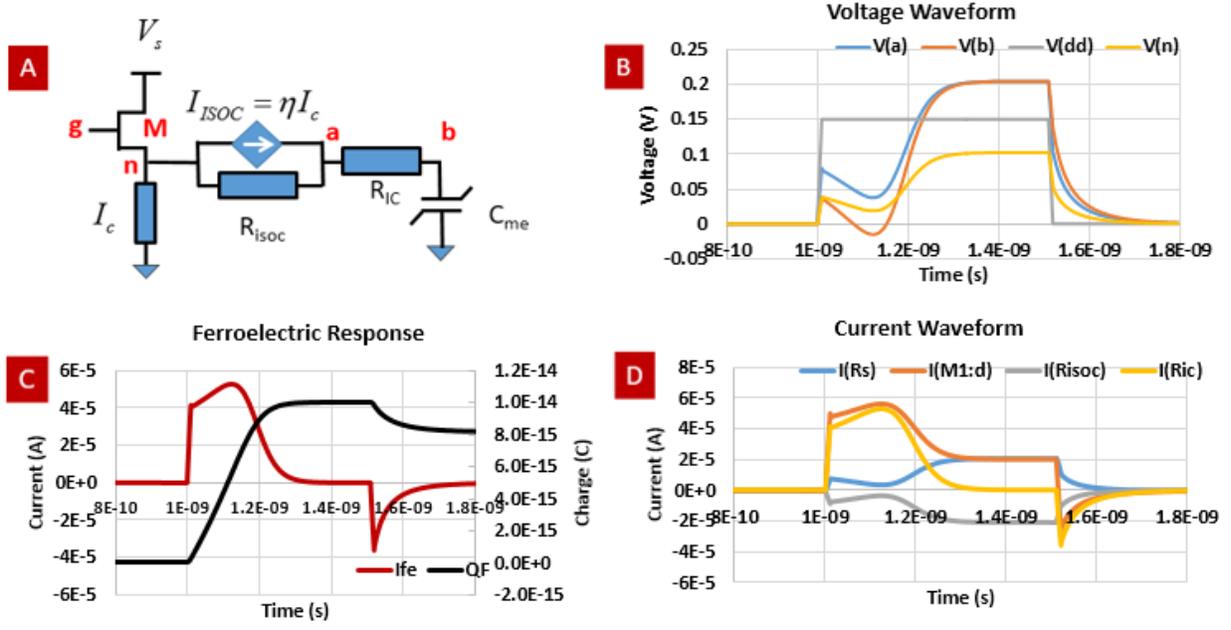

*Figure 4. SPICE Simulation of MESO showing the self-consistent charge dynamics of the MESO device **(for clarity of transient waveforms we simulated a 8 fC stored ferroelectric charge which is significantly higher than a scaled MESO logic device)** (A) Charge circuit model for MESO with ISOC current controlled current source and ME capacitor modeled with Landau-Khalatnikov equations B)Voltages applied/measured at supply transistor gate (**V(dd)**), magnetoelectric capacitor (**V(b)**), interconnect (**V(a)**) and the transistor terminal (**V(n)**) C) Current and charge across FE capacitor D) Currents measured through ISOC internal resistance ($R_{isoc}$), supply resistance ($R_s$), supply transistor (M) & interconnect resistance ($R_{IC}$) ( Pulse width=500ps, $R_{isoc}=R_s=5k$, Vdd=150mV (clk), Vg=1V (dc))*

We perform time domain self-consistent SPICE simulations comprehending the ISOC current controlled current source with the non-linear dynamics of the Ferroelectric. The typical switching dynamics are shown in Figure 4. The supply is turned on from 1 ns to 1.5 ns. The switching dynamics at node **b** (FE capacitor terminal) are consistent with a ferroelectric switching via internal polarization dynamics. The charge and current through the FE capacitor is shown in Figure 4.C. The current is consistent with classic FE switching pulse followed by a non-polarizing pulse at the end of the applied voltage.

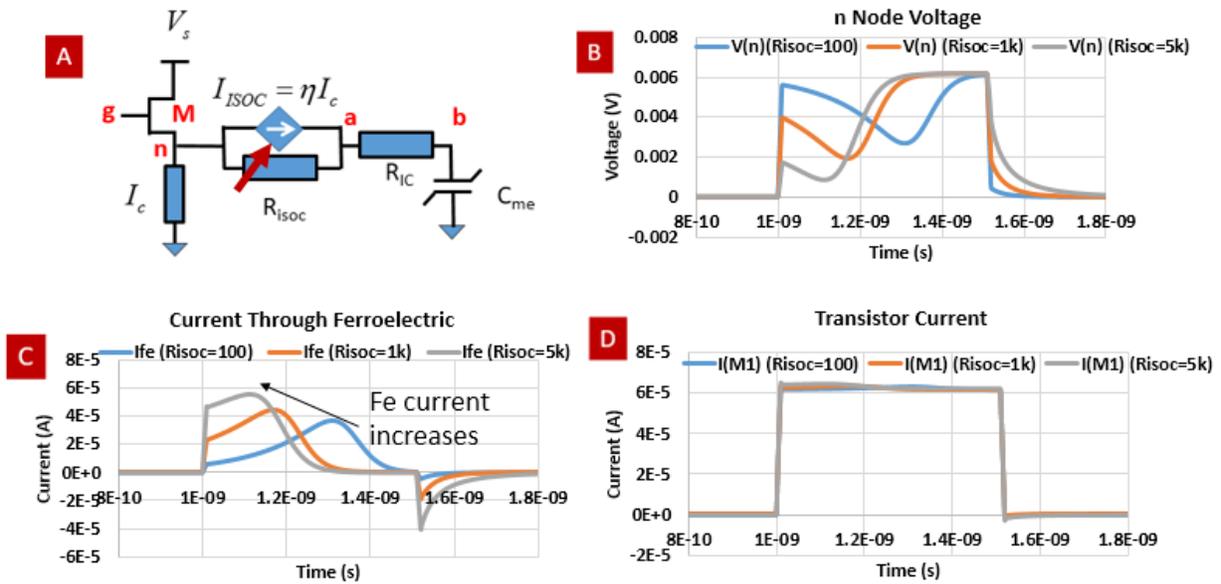

*Figure 5. **(Impact of ISOC internal current source resistance) (Impact of ISOC source resistance) (For clarity of transient waveforms we simulated an 8 fC stored ferroelectric charge which is significantly higher than a scaled MESO logic device)** SPICE Simulation of MESO showing the self-consistent charge dynamics of the MESO device B) Node voltage at n1 at varying values of $R_{isoc}$ (100 Ω, 1kΩ, 5kΩ) C) Ferroelectric node current (Sweep $R_{isoc}$, Pulse width=500ps, $R_s$=100, Vdd=150mV, Vg=1V). D) Transistor current at varying values of $R_{isoc}$ (100 Ω, 1kΩ, 5kΩ) For high ISOC resistance FE node current approaches the transistor current for high spin to charge conversion ratio.*

We show the impact of the ISOC internal resistance and in particular show that interconnect current is close to the supply current for high internal resistance SOC materials. The currents measured through the various branches show that ISOC generated charge current is shared between the internal shunt path resistance and the interconnect resistance. The extent of shunting via internal resistance depends on the resistivity of the ISOC material (Figure 5).

We study the impact of the supply resistor ($R_s$) on the switching dynamics of MESO. We show that the impact of supply path current (i.e., current through the resistor $R_s$) can be mitigated by a higher impedance without impacting the interconnect current and the switching dynamics of the ferroelectric (Figure 6).

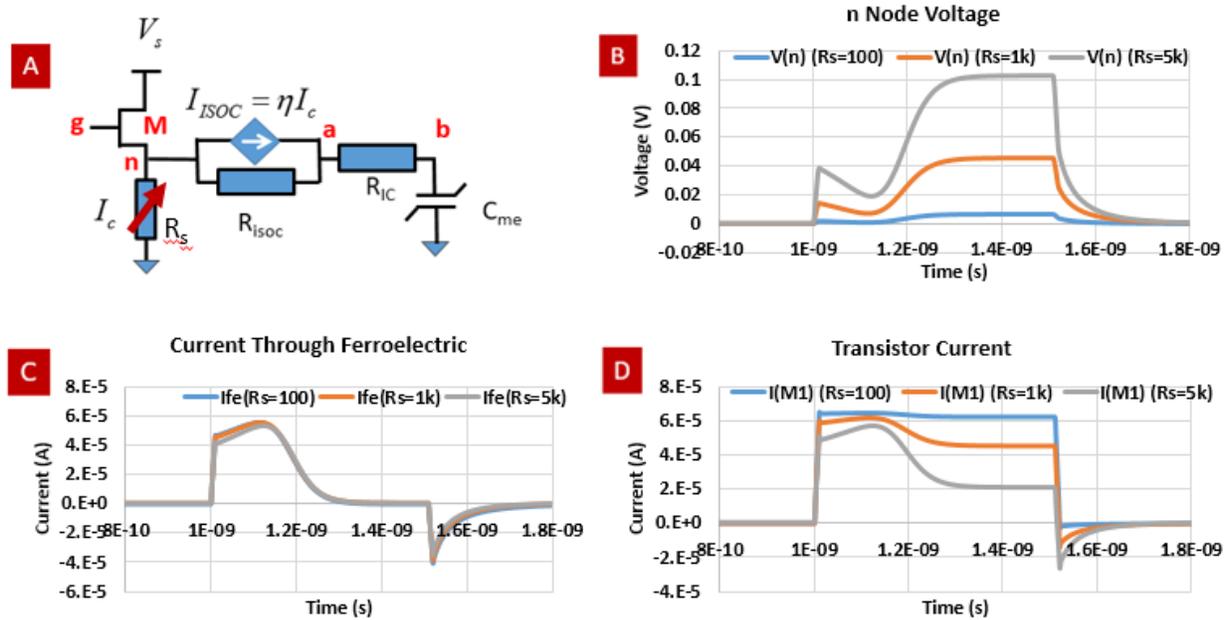

*Figure 6. **(Impact of $R_s$, Supply Resistance) (For clarity of transient waveforms we simulated an 8 fC stored ferroelectric charge which is significantly higher than a scaled MESO logic device)** A) SPICE Simulation of MESO showing the impact of $R_s$, supply resistance B) Node voltage at n at varying $R_s$ (100 Ω, 1kΩ, 5kΩ) C) Ferro-electric/Interconnect current through $R_{IC}$ D) Transistor*

current at varying $R_s$ (100 Ω, 1kΩ, 5kΩ) (Sweep $R_s$, Pulse width=500ps, Risoc=5k, Vdd=150mV (clk), Vg=1V(dc))

Scaled MESO switching operation comprehending all parasitic effects.

We further show the SPICE simulation for a sub 10 aJ switching operation of a MESO logic device comprehending: a) Energy stored in $C_{me}$ b) Energy dissipation in the Interconnect-$R_{IC}$; c) Energy dissipation in the Rashba source's resistance $R_{isoc}$; d) Supply resistor losses.

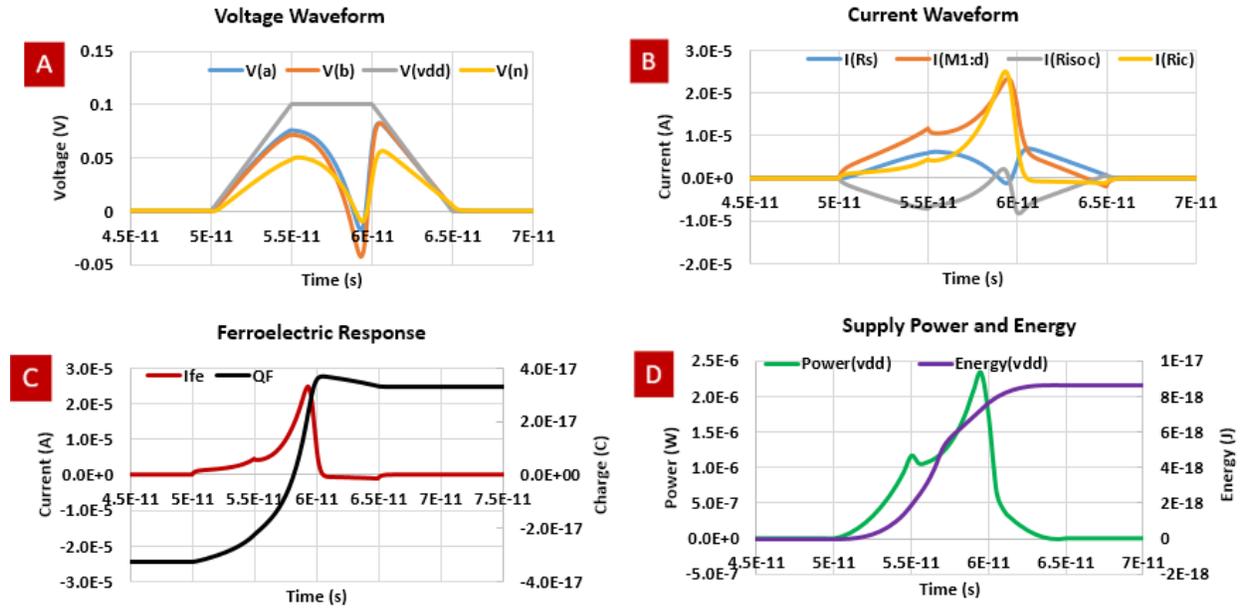

*Figure 7. **(Scaled MESO with 35 aC stored charge, sub 10 aJ switching energy)** A) Voltages applied/measured at supply transistor gate (**V(dd)**), magnetoelectric capacitor (**V(b)**), interconnect (**V(a)**) and the transistor terminal (**V(n)**) B) Currents measured through ISOC internal resistance ($R_{isoc}$), supply resistance ($R_s$), supply transistor (M) & interconnect resistance ($R_{IC}$) C) Ferroelectric node current and stored charge D) Supply power and energy vs time showing the total power delivered by the supply (Pulse width=15 ps, $R_{isoc}$=4 kΩ, $R_s$=8 kΩ, $R_{ic}$=1 kΩ, $V_{dd}$=100 mV (clk), $V_g$=0.8 V(dc))*

The total energy of MESO is smaller than 10 aJ for scaled device dimensions with scaled material properties. Figure 7. shows the charge dynamics of a scaled MESO with ~35 aC stored charge corresponding to 35 µC.cm$^{-2}$ FE polarization charge density on a 10 nm X 10 nm ME capacitor. The impact of the supply current path can be mitigated via use of high resistive path to limit the current to the required magnitude for switching (Figure 7.B). A supply voltage of 100 mV is applied for 15 ps in Figure 7A, and the resultant voltages at magnetoelectric capacitor (**V(b)**), interconnect (**V(a)**) and the transistor terminal (**V(n)**) are shown. Currents measured through ISOC internal resistance ($R_{isoc}$), supply resistance ($R_s$), supply transistor (M) & interconnect resistance (RIC) are shown in Figure 7B. FE polarization charge with retention and charge storage of ~ 35 aC can be observed in Figure 7C. The current in the interconnect follows the voltage difference across the transistor node and the FE node.

We probe the total power delivered by the supply for the scaled MESO to comprehend the effect of energy loss/storage from all the parts of the device. Figure.7D shows that the total integrated energy per switching transistion is ~ 10 aJ. We note that these devices assumed ~ 35 aC of stored polarization per device which is equivalent to 35 µC.cm$^{-2}$. Further significant scaling in total energy is possible with a reduction of the FE polarization.

## 5. Requirements for the resistivity of the spin-orbit coupling materials - Note on internal resistance of ISOC current controlled current source.

The output resistance of ISOC Current Controlled Current Source (CCCS) is obtained by dividing the open circuit voltage of the CCCS divided by the short circuit current [2, 3].The CCCS current source can also be converted to a current controlled voltage source (CCVS) by performing a Norton to Thevenin source conversion.

The ability of a current source to provide a current under resistive loading condition is improved as the internal resistance is increased. Research in spin-orbit coupling materials is opening up the possibility of high spin-orbit coupling materials with high intrinsic resistivity [8, 9], approaching 4-10 mΩ.cm. For example, a resistivity of 10 mΩ.cm [10] will provide an internal resistance of 5 kΩ - 20 kΩ for an ISOC spin to charge conversion layer with dimensions of 20 nm X 10-20 nm X 10-20 nm.

6. Misconceptions about MESO switching

Let us point out the differences between our analysis and the alternative treatment of switching performance in [6]. That work states:

*"the paper neglects a large energy cost associated with Ohmic dissipation that is unavoidable within the MESO scheme."*

In fact, the above derivation showed that we had accounted for Ohmic dissipation and this part of energy is not excessively large.

*"In estimating the total energy cost of this process, ref. [1] asserted that the dominant energy cost is associated with the electrostatic energy stored on the capacitor CME. ... Reference [1] did not take into account that, like any current source with an internal resistance much less than the load, more power will be dissipated within the internal resistance than delivered to the output, in this case much more."*

As shown in Section 2 below, in the case of the capacitor being charged via a current source with finite output resistance and a interconnect (with resistance), the energy dissipated in the resistances and the energy stored in the capacitor are closely related. Therefore we believe we have correctly accounted for the total power dissipation in this circuit.

*"However, by the nature of the materials that must be used within the spin-to-charge converter (a metal layer in contact to a material with strong spin-orbit coupling), it necessarily functions as a current source with a small internal resistance, RSCC."*

Please see section 5. A current source with small internal resistance will indeed interfere with the operation of MESO. We corrected the Figure in [1] to read $R_{isoc}$ = 10 k$\Omega$ (there was a mislabeling of the resistance magnitude value). The target material resistivity of the SOC materials is 1-10 m$\Omega$.cm. We further clarify the requirements of the spin to charge converter. We note that the material forming the current controlled current source is required to be a low conductivity material.

*"Therefore, for the full time that CME is biased with a steady voltage to drive magnetic switching (assumed in [1] to be a 100 mV bias for 100 ps), the voltage that is present across CME must also drop across the resistance RSCC, where it will generate a large current flow and a large amount of Ohmic dissipation. This current flow must circulate within the spin-to-charge converter for the full time required to drive magnetic switching; if the spin-to- charge converter is not energized, the magnetoelectric capacitor CME will discharge through RSCC and switching will not occur."*

The author of [6] has probably mistaken the time $\tau_{mag} \sim 200\,ps$ that it takes magnetization to switch for the time $\tau_{me} \approx 50\,ps$ it takes to charge the magnetoelectric capacitor. Once the capacitor is charged over time $\tau_{me}$, the switching of the multiferroic proceeds and then the exchange bias switches magnetization over time $\tau_{mag}$. However keeping the current source on for longer than $\tau_{me}$ it does not provide any benefit; and in fact, in our scheme the access transistor switches off the current in the supply path after this time and thus stops the ISOC current source. Therefore the author of [6] significantly over-estimated the on-time for the current source.

> *"Assuming the optimistic parameters stated in ref. [1]: that a voltage of 100 mV must be applied to the magnetoelectric capacitor for 100 ps to drive switching and that RSCC = 10 Ω, the energy cost due to Ohmic dissipation in RSCC for each switching event is (100 ps) (100 mV)2/(10 Ω) = 100 fJ."*

As we have pointed out above, the resistance in this path is much larger than $10\Omega$, i.e. $25k\Omega$. Together with the exaggerated time "on", this estimate is 4 orders of magnitude larger that the correct one, Eq. (2.7, 3.20). Also please refer to simulation of a scaled MESO in figure 7.

> *"There will be additional Ohmic dissipation due to current flow through the ferromagnetic injector, RFMI. Assuming, generously, that the spin-polarized charge current ($I_{in}$) is 100% polarized and that the spin-to-charge converter has unity efficiency, the magnitude of Iin must be at least approximately the same as the current generated within the spin-to-charge converter, which for the parameters stated above is ISCC = (100 mV)/(10 Ω) = 10 mA [2]. Using the value RFMI = 5 Ω projected in ref. [1], the minimum energy dissipated in $R_{FMI}$ during a 100 ps switching process is then (100 ps) (10 mA)2 (5Ω) = 50 fJ. Together, the sum of the Ohmic energy loss per switching event in RSCC and RFMI is therefore at least 150 fJ, the value I stated above."*

The author of [6] overlooks the dominant resistance in this path – presented by the access transistor. Its resistance combined with the wiring resistance is $R_{sum} = 78k\Omega$, which is much larger than the $5\Omega$ estimate used in [6]. This leads to a current in this path $I_{on} = 1.2\mu A$, which is much smaller than the estimate of 10 mA from [6]. Together with a shorter "on" time for the pulse we arrive at the estimate of energy lost in this path in Eq. (3.18), which is orders of magnitude smaller than in [6]. The total energy calculation presented in SPICE simulations (Figure 7) comprehends all the energy dissipation mechanisms.

## 7. Conclusions

In summary, we have presented here a detailed discussion on the energy consumption of a magnetoelectric spin-orbit logic gate, which uses magnetoelectric switching for switching the ferromagnet and a spin to charge conversion layer for generating the charge read out. We show analytically and with a compact circuit model and SPICE circuit simulation, the operating dynamics of MESO. We presented a time domain example where the total energy including all the dissipation methods is ~ 10 aJ per switching transition. Further significant scaling in total energy is possible with improvements in the materials properties of ISOC and magnetoelectrics.

**Acknowledgements:** We also acknowledge discussions and communications with Profs. Ramesh Ramamoorthy, Sayeef Salahuddin, Joerg Appenzeller, J.P. Wang, Nitin Samarth, Dr. Raseong Kim.

## 8. References

[1] Manipatruni, S., Nikonov, D.E. and Young, I.A., 2015. Spin-orbit logic with magnetoelectric nodes: A scalable charge mediated nonvolatile spintronic logic. Available online, ArXiv:1512.05428.

[2] Sedra, Adel S., and Kenneth Carless Smith. Microelectronic circuits. Vol. 1. New York: Oxford University Press, 1998.

[3] Kuo, Franklin. Network analysis and synthesis. John Wiley & Sons, 2006.

[4] E.-K. Tan, J. Osman, and D. Tilley, Physica Status Solidi (b) 228, 765 (2001); Tilley D R and Zeks B 1984 Solid State Commun. 49 823

[5] D. E. Nikonov and I. A. Young, "Benchmarking of Beyond-CMOS Exploratory Devices for Logic Integrated Circuits", IEEE J. Explor. Comput. Devices and Circuits 1, 3-11 (2015).

[6] D. C. Ralph, Comment on "Spin-Orbit Logic with Magnetoelectric Nodes: A Scalable Charge Mediated Nonvolatile Spintronic Logic" (arXiv: 1512.05428), available online: arXiv: 1607.06690.

[7] S. Rakheja and A. Naeemi, "Interconnects for Novel State Variables: Performance Modeling and Device and Circuit Implications", IEEE Trans. Electron Devices, v. 57, no. 10, p. 2711-2718 (2010).

[8] Karube, S., Kondou, K. and Otani, Y., 2016. Experimental observation of spin to charge current conversion at non-magnetic metal/Bi2O3 interfaces. arXiv preprint arXiv:1601.04292.

[9] Kushwaha, S.K., Pletikosić, I., Liang, T., Gyenis, A., Lapidus, S.H., Tian, Y., Zhao, H., Burch, K.S., Lin, J., Wang, W. and Ji, H., 2016. Sn-doped Bi1. 1Sb0. 9Te2S bulk crystal topological insulator with excellent properties. Nature communications, 7.

[10] Personal communication: Profs. J.P. Wang, Nitin Samarth